# Strong anisotropic influence of local-field effects on the dielectric response of α-MoO$_3$

*L. Lajaunie\*, F. Boucher, R. Dessapt, and P. Moreau.*

Institut des Matériaux Jean Rouxel, (IMN) – Université de Nantes, CNRS,  2 rue de la Houssinère - BP 32229, 44322 Nantes Cedex 3, France.

\* Corresponding author: luc.lajaunie@cnrs-imn.fr

**ABSTRACT**

Dielectric properties of α-MoO$_3$ are investigated by a combination of valence electron-energy loss spectroscopy and *ab initio* calculation at the random phase approximation level with the inclusion of local-field effects (LFE). A meticulous comparison between experimental and calculated spectra is performed in order to interpret calculated dielectric properties. The dielectric function of MoO$_3$ has been obtained along the three axes and the importance of LFE has been shown. In particular, taking into account LFE is shown to be essential to describe properly the intensity and position of the Mo-N$_{2,3}$ edges as well as the low energy part of the spectrum. A detailed study of the energy-loss function in connection with the dielectric response function also shows that the strong anisotropy of the energy-loss function of α-MoO$_3$ is driven by an anisotropic influence of LFE. These LFE significantly dampen a large peak in $\varepsilon_2$, but only along the [010] direction. Thanks to a detailed analysis at specific *k*-points of the orbitals involved in this transition, the origin of this peak has not only been evidenced but a connection between the inhomogeneity of the electron density and the anisotropic influence of local-field effects has also been established. More specifically, this anisotropy is governed by a strongly inhomogeneous spatial distribution of the empty states. This depletion of the empty states is localized around the terminal oxygens and accentuates the electron inhomogeneity.

**I. INTRODUCTION**

α-MoO$_3$ is the thermodynamically stable phase of the molybdenum trioxide in ambient conditions and is described in an orthorhombic unit cell (space group *Pbnm*) with the cell parameters *a* = 3.9624(1), *b* = 13.860(2), *c* = 3.6971(4) Å.[1] The structure consists in $^2/_\infty$[MoO$_3$] sheets that are orientated perpendicular to the [010] *y* axis, and are held together by van der Waals interactions (Fig. 1(a)). The notation $^n/_\infty$ is commonly used to indicate the dimensionality of the mineral block, $^1/_\infty$ and

$^2/_\infty$ referring to metal-oxide chains and layers, respectively. The $^2/_\infty$[MoO$_3$] blocks are built upon linear $^1/_\infty$[MoO$_4$] chains composed of distorted [MoO$_6$] octahedra sharing *cis*-equatorial edges (Fig. 1(b)). These chains are then connected via sharing apical corner to give rise to the layers. The distorted octahedra present three crystallographically inequivalent oxygen sites involving singly coordinated (terminal) oxygen, twofold coordinated oxygen and threefold coordinated oxygen.

In the last decades, MoO$_3$ has received considerable attention in a wide range of technological applications such as catalysis,[2,3] gas detection,[4] inorganic light emitting diode,[5] Li-ion batteries,[6] or electrochromic and photochromic devices.[7,8] In particular, there is a strong revival in the study of the α-MoO$_3$ phase, triggered by recent developments on its nanostructuration. A large variety of α-MoO$_3$ nanostructures can be synthesized including nanowires, nanobelts, nanolayers, nanorods, nanoparticules and mesoporous films.[9–12] Compared to the bulk form, these nanostructures present enhanced physical and chemical properties and are promising candidates for several applications including field-emitting devices,[11] positive materials for Li-ion batteries,[10,13,14] and electrochemical supercapacitors.[9] Furthermore, recent studies have highlighted the enhancement of the photoluminescence intensity of such nanostructures, opening a way for the development of advanced optoelectronic nanodevices.[12,15]

Since important progress has been obtained on the synthesis of molybdenum oxide nanostructures, the next challenge consists of the development of appropriate experimental tools for their characterization. Valence electron energy-loss spectroscopy (VEELS) performed in a transmission electron microscope is an unrivaled tool to probe the dielectric function (DF) at the nanoscale. Plasmonic properties of several nano-objects including silver nanoparticles, silver nanorods and gold nanodecahedra have been successfully determined through VEELS experiments.[16–18] Interpretation of VEELS spectra is however not straightforward and requires careful analysis. Several excitation processes and features including plasmons, interband transitions, surface effects and relativistic losses have to be taken into account before an interpretation of the VEELS spectra can be made.[19–22] The dielectric response of nanomaterials can also be determined through discrete dipole approximation, boundary element or effective-medium theory (EMT) approaches.[23–25] The EMT is particularly interesting as it is widely used for a large variety of heterostuctures and it can also include local-field effects (LFE).[26] Previous works have shown the good agreement between the results obtained from *ab initio* and EMT calculations.[27,28] All these methods require however the previous determination of the dielectric properties of the bulk material. Until now, even if the optical properties of α-MoO$_3$ have been intensively studied, they were only determined at low energy and within the *ac* plane.[29,30]

In this context, combination of first principle calculations from density functional theory (DFT) and VEELS experiments is a powerful strategy to derive the dielectric response of materials as well as to help to the interpretation of the VEELS features. This approach has been successfully applied to a large variety of materials.[31–34] The calculation of electronic excitations is however a highly non-trivial problem[21,35,36] and the DFT, being ground-state properties designed, can be insufficient to fully describe all VEELS features. More complete calculations are based on the resolution of the Bethe-Salpeter equation, which takes into account both excitonic effects and local-field effects but this method is very costly in terms of computational resources.[37] LFE arise in inhomogeneous electron systems, the individual dipoles responding to a local field induced by the other dipoles of the system and/or by an external field.[38,39] LFE alone, *i.e.* without including excitonic effects, can be computed at reasonable effort in the random phase approximation (RPA) by considering the reciprocal space range dependence of the microscopic dielectric matrices. Macroscopic quantities are obtained afterwards by including off-diagonal elements of theses matrices in the matrix inversion process. Several papers have already shown that LFE should be taken into account to properly describe optical and VEELS experiments.[33,40–43]

In order to compare calculations with experiments, several strategies can be chosen, depending on which levels the comparison is performed. For example, calculated real and imaginary parts of the DF can be compared to those derived from the experimental VEELS spectra via a Kramers-Kronig analysis (KKA).[44] This KKA approach however fails to take into account surface effects and Čerenkov losses as well as the intermixing, for anisotropic crystals, between the dielectric responses parallel and perpendicular to the electron beam.[22,45] Such effects can drastically hinder the determination of the dielectric properties. According to the dielectric function calculated in this paper, the acceleration voltage should be reduced below 40 kV to avoid any relativistic effects (leading to Čerenkov losses) in the low loss spectrum of α-$MoO_3$. Consequently, optical and dielectric properties of α-$MoO_3$ can not be derived though routinely performed KKA. The comparison can also be performed at the VEELS spectra level: the aforementioned effects can then be included in the calculations to generate calculated VEELS spectra and compared them to the experimental spectra deconvoluted from the multiple scatterings.[46] This last approach should be preferred when subtle changes in dielectric properties are expected.

In this paper, we use a combination of EELS experiments and density functional theory (DFT) calculations to investigate the dielectric properties of bulk α-$MoO_3$. LFE are introduced at the RPA level, and for the reasons explained above, the comparison between experiment and calculation is performed at the VEELS spectrum level. We are able to show that the energy-loss function of α-$MoO_3$ presents a strong anisotropy, which is strongly modulated by local-field effects arising from the inhomogeneous electron distribution. This work constitutes the basis for further studies on more complex structures and nanostructures

This paper is organized as follows. Experimental setup and calculation details are introduced in sections II and III, respectively. The result of the calculated static dielectric constant is presented in section IV together with the comparison of simulated and experimental VEELS spectra. Finally a detailed study of the influence of local-field effects on the energy-loss and macroscopic dielectric functions is discussed in section V.

## II. EXPERIMENTAL DETAILS

Commercial molybdenum trioxide (α-MoO$_3$) powder (Acros, 99%) was crushed in an agate mortar with alcohol, then dispersed by ultrasound and finally deposited onto a holey carbon grid. Electron energy-loss spectra were acquired using a Hitachi HF2000 transmission electron microscope (100 kV) equipped with a cold field emission gun and a modified Gatan PEELS 666 spectrometer. The energy resolution, measured as the full width at half maximum of the zero loss peak (ZLP) was 0.8 eV and the energy dispersion was 0.10 eV/pixel. Convergence and collection angle were 1.4 and 9 mrad, respectively. Experiments were performed at liquid nitrogen temperature to minimize carbon contamination and electron beam damage, which are known to be important in α-MoO$_3$ EELS experiments.[47,48] The orientation of the probed crystal was obtained by electron diffraction prior to spectrum recording. All spectra were first gain and dark count corrected and then deconvoluted by the ZLP using the PEELS program.[49] Next, the single scattering spectra were obtained following Stephen's procedure,[50] which also gave the $t/\lambda$ ratio ($t$ the thickness of the analyzed crystal and $\lambda$ the inelastic mean free path). In order to obtain intensities as probability/eV units and to get an absolute comparison with calculations, the experimental spectra were divided by the zero-loss intensity integrated between -3 and 2 eV.[51]

## III. COMPUTATIONAL DETAILS

The ground state electronic structure of α-MoO$_3$ was obtained within the DFT formalism using the Perdew-Burke-Ernzerhof parameterization of the generalized gradient approximation (PBE-GGA ).[52] Calculations have been carried out with the *ab initio* total energy and molecular dynamics programme VASP[53] (Vienna ab initio simulation package). Projector augmented-wave (PAW) pseudopotentials were used.[54,55] Standard PAW was chosen for O (PAW PBE O: 2s$^2$2p$^4$) and semicore states were included for Mo (PAW PBE Mo_sv: 4s$^2$4p$^6$5s$^1$4d$^5$). Initial structural data were taken from the work of Sitepu.[1] Atomic positions were optimized by minimizing the residual Hellmann-Feynman forces on the atoms. Since van der Waals interactions are not taken into account on traditional DFT/GGA calculations, the lattice parameters were kept constant. Complete structural optimizations taking into account van der Waals contacts were also performed by using the semi-empirical DFT-D2 method implemented in VASP.[56] Results of the two structural methods and a detailed comparison with previous experimental and theoretical works can be found in the Supplemental Material.[1,57–59] However, whatever the structural optimization approach used, the dielectric properties up to 80 eV were found to be identical. Structural parameters obtained from the first method have thus been used in the following. The self-consistency on electronic density was obtained with a 600 eV plane wave energy cut-off and a (8×2×8) Monkhorst-Pack *k*-point mesh (16 *k*-points in the irreducible part of the Brillouin zone (IBZ)). Forces on atoms were minimized down to 0.02 eV/Å for the structural relaxation.

The frequency dependent DF including or not LFE were obtained by using the GW[60] and linear optic[61] routines implemented in the VASP 5 version, respectively. The microscopic DF was calculated at the RPA level by evaluating the microscopic polarizability matrices at $q \to 0$. A plane wave energy cut-off of 450 eV was found large enough to get converged quantities. For the optic routines, a (9×2×10) *k*-point grid was used (60 *k*-points in the IBZ). The frequency dependence was evaluated up to 80 eV, Kohn-Sham wave functions being evaluated for many additional empty bands (336 bands above the Fermi level). The results previously obtained from the WIEN2k/LAPW method were taken as reference in order to check on the accuracy of the pseudopotentials. For the GW routines, a randomly shifted[62] (12×4×12) k-point grid was used corresponding to a total of 576 k-points in the IBZ. The calculation of the quasiparticle energies was bypassed and the macroscopic DF including LFE at the RPA level was obtained from the inversion of the full microscopic dielectric tensors, taking into account the range dependencies with respect to reciprocal lattice **G** vectors.[37,61] Converged LFE were obtained with a set of **G** vectors defined by a plane wave energy cut-off of 75 eV, *i.e.* 318 **G** vectors.

In order to allow for absolute comparison, the dynamic DF calculated by VASP were inserted in previously published formulae taking into account the geometry of the EELS experiments together with surface and relativistic effects.[46] The sample thickness, the only parameter needed in the aforementioned formulae, was determined by estimating the inelastic mean free path using the modified Iakoubovskii formula.[44,63] A 0.8 eV Gaussian broadening (experimental energy resolution) was subsequently applied to the theoretically generated VEELS spectrum.

## IV. RESULTS

### A. STATIC DIELECTRIC CONSTANT

The static dielectric constants calculated, with and without local-field effects, along the three axes and the corresponding averaged values are given in Tab. I. The following sum rule for the static dielectric constant can be obtained from the first Kramers-Kronig dispersion relation:[64]

$$\varepsilon_1(0) - 1 = \frac{2}{\pi} \int_0^\infty \frac{\varepsilon_2(\omega)}{\omega} d\omega \qquad (1)$$

Equation (1) states that the static dielectric constant is a measure of the weighted sum of all absorption processes. Inclusion of LFE leads to a decrease of the static dielectric constant as already reported for other materials[38,40,41]. In particular the averaged value of $\varepsilon_1(0)$ decreased from 7.0 to 5.3 in closer agreement to the experimental value of 5.7 determined on polycrystalline α-MoO$_3$ thin films by Deb et al.[29] This result suggests that excitonic effects should have a weak influence on the sum of all the absorption processes of α-MoO$_3$. In addition LFE have a stronger influence along the x and y axes than along the z axis where the static dielectric constant is barely modified. LFE lead thus to a decrease of the integrated intensity of $\varepsilon_2(\omega)$ along the x and y axes whereas LFE should have a minor influence along the z axis. This will be confirmed when analyzing the DF in Section V. The static dielectric constant gives thus a first hint on the importance of LFE at a lesser computational effort than the one required for the calculation of the frequency dependent DF.

### B. DIRECT COMPARISON THEORY/VEELS EXPERIMENT

When dealing with an uniaxial anisotropic crystal, the whole dielectric tensor can be approximated by a two components one that contains only $\varepsilon_{\parallel}(\mathbf{0},\omega)$ and $\varepsilon_{\perp}(\mathbf{0},\omega)$. These are defined according to directions that are parallel and perpendicular to the anisotropic crystal axis, respectively, and are the parameters used in our formula for calculating the full VEELS spectra.[46] Considering the two dimensional nature of the α-MoO$_3$ crystal structure, this two components simplification can be applied. $\varepsilon_{\parallel}(\mathbf{0},\omega)$ and $\varepsilon_{\perp}(\mathbf{0},\omega)$ are referring in that case to the calculated $\varepsilon_y(\mathbf{0},\omega)$ and $\varepsilon_{x,z}(\mathbf{0},\omega)$ quantities, respectively, where $\varepsilon_{x,z}(\mathbf{0},\omega)$ is the average of the tensor values calculated for $\varepsilon_x(\mathbf{0},\omega)$ and $\varepsilon_z(\mathbf{0},\omega)$. The experimental spectrum having been recorded along the [210] zone axis, both $\varepsilon_{\parallel}(\mathbf{0},\omega)$ and $\varepsilon_{\perp}(\mathbf{0},\omega)$ components are thus probed by the VEELS experiment due to the oblique incidence of the electron beam with respect to the anisotropic axis direction. The angle between the [210] zone axis and the axial [010] direction being equal to 30°, the intermix between the two components is expected to be small, with a predominant contribution coming from $\varepsilon_\perp(\mathbf{0},\omega)$. Due to preferential orientation growth, all probed α-MoO$_3$ platelets were effectively orientated close to this zone axis thus preventing the easy recording of spectra with a stronger $\varepsilon_\parallel(\mathbf{0},\omega)$ magnitude.

Comparison between experimental and theoretical spectra (with and without the inclusion of LFE) is given in Fig. 2. An excellent overall agreement is observed between the experimental and calculated spectra including LFE. In particular, this inclusion is stressed as essential to describe properly the intensity and position of the Mo-N$_{2,3}$ edges resulting from Mo-$4p$ excitation. The maximum energy of the Mo-N$_{2,3}$ edges is calculated around 40.2 and 47.8 eV, without and with LFE respectively, whereas the experimental one is at 49.6 eV. Local-field effects lead thus to a shift to higher energy and a decrease in intensity for the Mo-N$_{2,3}$ edges as already observed for other semicore states.[38,42,62,65] Focusing now on the low-energy part of the spectrum (E<22 eV), LFE induce an increase of intensity in the region situated around 15-20 eV, thus improving the agreement between experimental and calculated spectra. It is worth noting that thanks to an analogous band to band analysis as that described in section V, we are able to show that this increase of intensity results from the influence of LFE on the transitions from the O-$s$ states towards the $t_{2g}$ manifold. Local-field effects may thus explain the lack of relative intensity already observed on calculated VEELS spectra without LFE of similar compounds such as TiO$_2$ and SrTiO$_3$.[32,66] Slight differences are however observed between calculations and experimental spectra concerning the energy position of both A and B structures that are underestimated by 1 eV. This may be ascribed to the underestimation of the band gap by DFT calculations. Calculations also overestimate by roughly 10% the absolute intensity when compared to the experimental spectra. This difference is ascribed to the overestimation of the inelastic mean free path by the modified Iakoubovskii, formula which thus leads to an increase of the absolute intensity.[44] In addition, the calculated spectra present more resolved features when compared to the experimental one. Such mismatch may be ascribed to the **q** dependence of the dielectric function and to the lifetime of the excitation process which are not taken into account in the present calculations. An increased broadening of the VEELS feature together with a stronger influence of LFE is expected with increasing **q**.[42,67] Despite these slight differences, the overall agreement between the experimental spectra and energy-loss function simulation including LFE confirms our calculations of the dielectric tensor properties and allow us to use them for further investigations.

### C. ANISOTROPY OF THE ENERGY-LOSS FUNCTION

Comparison of calculated energy-loss functions ($Im[-1/\varepsilon]=\varepsilon_2/(\varepsilon_1^2+\varepsilon_2^2)$), including or not LFE along the three axes is given in Fig. 3. Focusing first on the high energy part of the spectra (E>22 eV), the influence of LFE is clearly seen on the shape and intensity of the spectra in a similar manner for the three directions. As already observed for Ni and Li related compounds, the strength of LFE on semicore edges is related to the localization of the initial and final electron wave functions involved in the considered transition.[38,68] Due to the atomic nature of these LFE, *i.e.* short range dependence in real space, a high number of **G** vector must be included (318 **G** vectors, *i.e.* long range dependence in reciprocal space). Thus, these atomic scale LFE are barely influenced by the orientation of the crystal as already observed for Ti-$M_{2,3}$ edges.[33] For the low-energy part of the spectra (E < 22 eV), due to the more delocalized nature of the levels involved in the transitions, about 100 **G** vectors have to be included in the calculation of the macroscopic dielectric function for the low energy part of the spectra. These LFE are therefore less localized in real space than the one influencing the Mo-$N_{2,3}$ edges. A rough estimate of the spatial extension of the LFE, based on the number of G vectors necessary for describing properly the excitation processes, gives for the low energy part an inhomogeneity at a scale larger than 3 Å, whereas for the Mo-$N_{2,3}$ edge one has to refine the description of this inhomogeneity down to a scale as small as 1 Å. Minor modification of the energy-loss functions are observed along the *x* and *z* axes: LFE induced a small increase of the intensity in the 15-20 eV energy range while the maximum of the B structure is shifted by nearly 1 eV towards high energy. On the contrary, LFE drastically modify the energy-loss function calculated along the *y* direction. Without LFE, it presents similar features to those calculated along the *x* and *z* axes especially with a large peak (B) calculated at 13 eV. Once LFE are included, the B structure disappears and is replaced by a succession of small intensity peaks; the spectral weight being shifted towards higher energy and thus leading to a strong anisotropy of the energy-loss function. In the next section, we will focus our investigations on the low-energy part of the dielectric response to have a better insight onto the anisotropy of the energy-loss function. For this purpose, an examination of the dielectric function is required.

### V. DISCUSSION

#### A. ANISOTROPIC INFLUENCE OF LOCAL-FIELD EFFECTS

Figs. 4 shows the real and imaginary parts of $\varepsilon(\mathbf{0},\omega)$ and the associated loss functions calculated, without and with LFE, along the *x* axis. Without inclusion of LFE, the B structure in the loss function occurs at the point where $\varepsilon_1$ crosses zero with a positive slope (~11 eV). By analogy with a free electron gas system, one could characterize this structure as a plasmon, a collective excitation mode.[21] The physical interpretation of the A structure is less straightforward. It could result either from a plasmonic behavior (around 5.5 eV) or from interband transitions (strong absorption peak in $\varepsilon_2$ around 4 eV). Even if the low-energy part (E < 8 eV) of the loss function seems barely sensitive to local-field effects, these latter drastically modify the dielectric function as shown in Fig. 4(b). LFE strongly dampen the intensity of the sharp peak observed around 4 eV in the imaginary part of the dielectric function (intensity of 17.1 and 6.7 without and with LFE, respectively). Consequently, the real part does not cross zero anymore around the A structure. From a macroscopic point of view, this structure has thus lost its possible collective excitation origin. This shows that even if the loss function seems barely affected by LFE, these latter have to be introduced to avoid any misinterpretation of the dielectric function.

Real and imaginary parts of $\varepsilon(\mathbf{0},\omega)$ calculated without and with LFE along the $y$ axis and the associated loss functions are displayed in Fig. 5. In this orientation and without LFE included, $\varepsilon_2$ shows two strong absorption peaks located around 4 and 7 eV, the intensity of the 7 eV peak being twice as strong as that along the $x$ axis. Furthermore the B structure in the loss function clearly results from a plasmonic behavior (zero crossing of $\varepsilon_1$ with a positive slope around 12 eV) with large plasmon strength. The plasmon strength is defined, in the single pole plasmon approximation, as the energy difference between the two points where $\varepsilon_1$ crosses zero with a negative and positive slope.[69] Once LFE are included, the intensity of the 4 and 7 eV absorption peaks are strongly dampened. The whole behavior of $\varepsilon_1$ at higher energy is consequently strongly modified: the plasmon strength is strongly reduced and the B structure is now replaced by small intensity peaks coming from both plasmonic and single excitations. Since the damping of the 4 eV absorption peak is also observed along $x$ and since no change occurs on the B structure along this axis, the whole anisotropy of the energy-loss function at low-energy lies in the damping of the 7 eV absorption peak along the $y$ axis. It is thus of primary importance for the purpose of this paper to understand the anisotropic influence of local-field effects on this peak.

The dielectric response along the $z$ axis and the associated loss-functions are displayed in the Supplemental Material. In this orientation, the dielectric functions together with the loss function are barely modified by the inclusion of LFE thus showing that the electron system is more homogenous along this direction than along the others. This, conjugated with the strong decrease of $\varepsilon_2$ induced by LFE at low energies along the *x* and *y* axes, explains the whole behavior of the static dielectric constant observed in the Section IV. The decrease of the static dielectric constant is thus triggered by the influence of LFE at low energy (E < 8 eV).

**B. INTERPRETATION OF THE 7 eV ABSORPTION PEAK AND RELATION TO THE SPATIAL INHOMOGENEITY OF THE EMPTY STATES**

The bands involved in the considered peak can be found by carefully analyzing the matrix elements for each band combination by using the OPTIC program of Wien2k. In order to shed light onto the origin of the 7 eV absorption peak, Fig. 6 shows the imaginary part of $\varepsilon(\mathbf{0},\omega)$ calculated along the *y* axis by considering only the transitions from the bands 53-56 bands towards the 77-78$^{th}$ bands. It is clear that $\varepsilon_2(\mathbf{0},\omega)$, along the *y* axis, is dominated around 7 eV by transitions involving the aforementioned bands. These bands are highlighted onto the band structure plot of α-MoO$_3$ shown in Fig. 7. This band structure is almost identical to that published and extensively discussed by Scanlon *et al.*[58] The bands 53 to 56 lie at the top of the valence band and show a predominant non bonding O-$p_y$ character along the high-symmetry points while the bands 77-78 lie in the $e_g$ manifold. It corresponds to a ligand to metal charge transfer (LMCT): an electronic excitation from ligand non bonding states to the metal *d* band presenting σ* type Mo-$d(e_g)$/O-*p* interactions. To complete the interpretation of the 7 eV peak, we have isolated and analyzed the "53-56 to 77-78" matrix elements for each *k*-points of the full list used for the calculation of the dielectric tensor (60 *k*-points in total). The *k*-points giving rise to the 7 eV peak do not lie along the high-symmetry points but close to three *k*-points, labelled L$_1$, L$_2$ and L$_3$ in the following, whose respective coordinates are (0;0.33;0.19), (0;0.33;0.45) and (0.1;0.33;0). The atomic orbital decompositions of the bands 53-56 and 77-78 at the $L_1$ point are illustrated in Fig. 8. For the valence bands, the character of the wave function is clearly oxygen rather than molybdenum (89% and 11% respectively) with a clear prevalence of the O-$p$ character (88% of the total states). A similar trend is observed for the two other $k$-points, $L_2$ and $L_3$. The LMCT excitation is governed by the dipolar approximation ($\Delta l = \pm 1$) which also implies a spatial overlap between the initial and final states. Since the bands 53-56 are non bonding, from symmetry consideration, no overlap could exist between the O-$p$ states of the bands 53-56 and the Mo-$d(e_g)$ levels. Thus, due to the highly localized and atomic like O-$p$ character observed in the bands 53-56, the transition towards the bands 77-78 can only be observed if some atomic-like O-$s$ character is present in the conduction band. In other words, the vertical transition from 53-56 towards 77-78 will only be present if the $k$-point symmetry allows a contribution of O-$s$ levels in the bands 77-78. The O-$s$ states being isotropic in nature, the spatial directionality of the 7 eV transition is therefore only given by the initial states. The stronger intensity of the considered peak along the $y$ axis rather than along the other directions can thus be attributed to the predominant O-$p_y$ character of the bands 53-56. Fig. 9 shows an electron density like representation (square modulus of the wave function) at the $k$-point $L_1$ for the 53-56$^{th}$ and 77-78$^{th}$ bands. While the three oxygen sites contribute to the electron density for the valence band, missing contribution are clearly highlighted around the terminal oxygens for the conduction band. The O-$s$ character contribution of the terminal oxygens represents less than 4% of the total O-$s$ states in the conduction band; they have thus a minor role in the 7 eV transition. Furthermore, the total amount of the O-$p_y$ character for the bands 53-56 concerning the doubly and threefold coordinated oxygen represents 24% and 11% of the total states at the $L_1$ and $\Gamma$ points, respectively. This explains thus the weaker intensity of the 7 eV absorption peak observed along the high-symmetry points. The depletion of the O-$s$ character around the terminal oxygens strengthen the inhomogeneity of the electron density along the *y* axis and explain thus the stronger influence of the LFE along this direction for the 7 eV peak. To check the influence of the electron density deficiency, we have performed analogous calculations on a modified structure of α-MoO$_3$ where the spacing between the $^2/_\infty$[MoO$_3$] sheets has been increased by 2 Å. The ratio of the lattice parameters *b/a* is thus increased from 3.5 to 4.5 for the original and modified structure, respectively. The imaginary part of $\varepsilon_2(\mathbf{0},\omega)$ calculated for the two structures along the *x* and *y* axes, with and without LFE, is shown in Fig. 10. In the absence of LFE and whatever the crystallographic orientation, $\varepsilon_2(\mathbf{0},\omega)$ shows only minor changes with the $^2/_\infty$[MoO$_3$] sheets spacing. These small modifications can be mainly attributed to volume change.[70] This confirms the weak electronic interaction between the $^2/_\infty$[MoO$_3$] sheets. When LFE are included, $\varepsilon_2(\mathbf{0},\omega)$ calculated along the *y* axis is heavily influenced by the sheets spacing whereas minor changes are only observed along the *x* axis. The increase of the sheet spacing by only 2 Å leads thus to an increase of the inhomogeneity of the electron density significantly enough to strongly dampened the variation of $\varepsilon_2(\mathbf{0},\omega)$ in the low energy part the spectrum. We would like to emphasize the importance of these results. Even if anisotropic influences of LFE on the dielectric function were already observed in layered materials[33,71], in this work clear connection has been made between the anisotropy of LFE and an inhomogeneous spatial distribution of the empty states.

## VI. CONLUSION

The dielectric response of α-MoO$_3$ has been studied by a thorough comparison of *ab initio* calculation based on the density functional theory and valence electron energy-loss spectroscopy. Thanks to this combined investigation, the dielectric properties along the three axes have been determined and the importance of local-field effects has been clearly highlighted. This should have cleared the path for the future works focusing on the dielectric properties of nanostructures based on α-MoO$_3$. From a more fundamental point of view, we have shown that these LFE exert an anisotropic influence on the dielectric function leading to a strong anisotropy of the energy-loss function. Specifically, LFE strongly dampen an absorption peak located near 7 eV but only along the *y* axis. We have shown that this peak results from transitions from the O-*p* states towards the O-*s* states. By a meticulous analysis at the specific *k*-points giving rise to this absorption peak, we have observed a strongly inhomogeneous spatial distribution of the empty states and a strong depletion of the empty states around the terminal oxygens. This depletion leads thus to a strengthening of the electron density inhomogeneity along the *y* axis. Thanks to this observation, we have correlated the anisotropic influence of LFE with the anisotropy of the empty states.

These results, in addition to providing more insight onto the dielectric response of α-MoO$_3$, show the intimate connection between local-field effects and the layered nature of the molybdenum trioxide. They can be thus extrapolated to a large variety of materials presenting a similar structure. In particular, our results suggest that the dielectric response of materials based on $^2/_\infty$[MoO$_3$] sheets, such as hybrid organic-inorganic molybdates[72,73], can be tailored along one direction by finely tuning the sheets spacing.

**REFERENCES**

[1] H. Sitepu, Powder Diffr. **24**, 315 (2009).

[2] B. Grzybowska-Swierkosz, Top. Catal. **11-12**, 23 (2000).

[3] T. Ressler, J. Wienold, R. Jentoft, and T. Neisius, J. Catal. **210**, 67 (2002).

[4] S. Barazzouk, R.P. Tandon, and S. Hotchandani, Sensors Actuat. B-Chem. **119**, 691 (2006).

[5] P.-S. Wang, I.-W. Wu, W.-H. Tseng, M.-H. Chen, and C.-I. Wu, Appl. Phys. Lett. **98**, 173302 (2011).

[6] C.V. Ramana, V.V. Atuchin, H. Groult, and C.M. Julien, J. Vac. Sci. Technol., A **30**, 04D105 (2012).

[7] T. He and J. Yao, J. Photoch. Photobio. C **4**, 125 (2003).

[8] J. Yao, K. Hashimoto, and A. Fujishima, Nature **355**, 624 (1992).

[9] T. Brezesinski, J. Wang, S.H. Tolbert, and B. Dunn, Nat. Mater. **9**, 146 (2010).

[10] J.S. Chen, Y.L. Cheah, S. Madhavi, and X.W. Lou, J. Phys. Chem. C **114**, 8675 (2010).

[11] L. Mai, F. Yang, Y. Zhao, X. Xu, L. Xu, B. Hu, Y. Luo, and H. Liu, Mater. Today **14**, 346 (2011).

[12] L.X. Song, J. Xia, Z. Dang, J. Yang, L.B. Wang, and J. Chen, CrystEngComm **14**, 2675 (2012).

[13] S. Berthumeyrie, J.-C. Badot, J.-P. Pereira-Ramos, O. Dubrunfaut, S. Bach, and P. Vermaut, J. Phys. Chem. C **114**, 19803 (2010).

[14] B. Gao, H. Fan, and X. Zhang, J. Phys. Chem. Solids **73**, 423 (2012).

[15] I. Navas, R. Vinodkumar, and V.P.M. Pillai, Appl. Phys. A: Mater. Sci. Process. **103**, 373 (2011).

[16] B.S. Guiton, V. Iberi, S. Li, D.N. Leonard, C.M. Parish, P.G. Kotula, M. Varela, G.C. Schatz, S.J. Pennycook, and J.P. Camden, Nano Lett. **11**, 3482 (2011).

[17] V. Myroshnychenko, J. Nelayah, G. Adamo, N. Geuquet, J. Rodr guez-Fernández, I. Pastoriza-Santos, K.F. MacDonald, L. Henrard, L.M. Liz-Marza n, N.I. Zheludev, and others, Nano Lett. **12**, 4172 (2012).

[18] J.A. Scholl, A.L. Koh, and J.A. Dionne, Nature **483**, 421 (2012).

[19] R. Erni and N.D. Browning, Ultramicroscopy **108**, 84 (2008).

[20] F.J. Garcia de Abajo, Rev. Mod. Phys. **82**, 209 (2010).

[21] V. Keast, J. Electron Spectrosc. Relat. Phenom. **143**, 97 (2005).

[22] M. Stöger-Pollach and T. Galek, Micron **37**, 748 (2006).

[23] F. de Abajo and A. Howie, Phys. Rev. B **65**, 115418 (2002).

[24] C. Brosseau, J. Phys. D: Appl. Phys. **39**, 1277 (2006).

[25] N. Geuquet and L. Henrard, Ultramicroscopy **110**, 1075 (2010).

[26] D. Aspnes, Am. J. Phys. **50**, 704 (1982).

[27] S. Botti, N. Vast, L. Reining, V. Olevano, and L.C. Andreani, Phys. Rev. B **70**, 045301 (2004).

[28] H.-C. Weissker, J. Furthmüller, and F. Bechstedt, Phys. Rev. B **67**, 165322 (2003).

[29] S. Deb and C. JA, J. Appl. Phys. **37**, 4818 (1966).

[30] M. Itoh, K. Hayakawa, and S. Oishi, J. Phys.: Condens. Matter **13**, 6853 (2001).

[31] M.K. Kinyanjui, P. Axmann, M. Wohlfahrt-Mehrens, P. Moreau, F. Boucher, and U. Kaiser, J. Phys.: Condens. Matter **22**, 275501 (2010).

[32] M. Launay, F. Boucher, and P. Moreau, Phys. Rev. B **69**, 035101 (2004).

[33] V. Mauchamp, G. Hug, M. Bugnet, T. Cabioc'h, and M. Jaouen, Phys. Rev. B **81**, 035109 (2010).

[34] O. Prytz, O.M. Lovvik, and J. Tafto, Phys. Rev. B **74**, 245109 (2006).

[35] C. Hébert, Micron **38**, 12 (2007).

[36] V. Keast, Micron **44**, 93 (2013).

[37] G. Onida, L. Reining, and A. Rubio, Rev. Mod. Phys. **74**, 601 (2002).

[38] F. Aryasetiawan, O. Gunnarsson, M. Knupfer, and J. Fink, Phys. Rev. B **50**, 7311 (1994).

[39] M. Fox, *Optical Properties of Solids* (Oxford Univ. Press, 2001).

[40] B. Arnaud and M. Alouani, Phys. Rev. B **63**, 085208 (2001).

[41] S. Baroni and R. Resta, Phys. Rev. B **33**, 7017 (1986).

[42] N. Vast, L. Reining, V. Olevano, P. Schattschneider, and B. Jouffrey, Phys. Rev. Lett. **88**, 037601 (2002).

[43] S. Waidmann, M. Knupfer, B. Arnold, J. Fink, A. Fleszar, and W. Hanke, Phys. Rev. B **61**, 10149 (2000).

[44] R.F. Egerton, *Electron Energy-loss Spectroscopy in the Electron Microscope* (Springer Verlag, 2011).

[45] J. Park and M. Yang, Micron **40**, 365 (2009).

[46] P. Moreau and M. Cheynet, Ultramicroscopy **94**, 293 (2003).

[47] D.E. Diaz-Droguett, A. Zuniga, G. Solorzano, and V.M. Fuenzalida, J. Nanopart. Res. **14**, 679 (2012).

[48] D. Wang, D.S. Su, and R. Schlogl, Z. Anorg. Allg. Chem. **630**, 1007 (2004).

[49] P. Fallon and C.A. Walsh, PEELS Program, University of Cambridge, England (1996).

[50] A.P. Stephen, PhD. Thesis, University of Cambridge, England, 1980.

[51] T. Stockli, J. Bonard, A. Chatelain, Z. Wang, and P. Stadelmann, Phys. Rev. B **61**, 5751 (2000).

[52] J. Perdew, K. Burke, and M. Ernzerhof, Phys. Rev. Lett. **77**, 3865 (1996).

[53] G. Kresse and J. Furthmuller, Phys. Rev. B **54**, 11169 (1996).

[54] P.E. Blöchl, Phys. Rev. B **50**, 17953 (1994).

[55] G. Kresse and D. Joubert, Phys. Rev. B **59**, 1758 (1999).

[56] S. Grimme, J. Comput. Chem. **27**, 1787 (2006).

[57] H. Ding, K.G. Ray, V. Ozolins, and M. Asta, Phys. Rev. B **85**, 012104 (2012).

[58] D.O. Scanlon, G.W. Watson, D.J. Payne, G.R. Atkinson, R.G. Egdell, and D.S.L. Law, J. Phys. Chem. C **114**, 4636 (2010).

[59] *See Supplemental Material at [URL Will Be Inserted by Publisher] for More Details About the Structural Optimization.* (n.d.).

[60] L. Hedin, Phys. Rev. **139**, A796+ (1965).

[61] M. Gajdos, K. Hummer, G. Kresse, J. Furthmuller, and F. Bechstedt, Phys. Rev. B **73**, 045112 (2006).

[62] P. Moreau and F. Boucher, Micron **43**, 16 (2012).

[63] K. Iakoubovskii, K. Mitsuishi, Y. Nakayama, and K. Furuya, Microsc. Res. Tech. **71**, 626 (2008).

[64] M. Dressel and G. Grüner, *Electrodynamics of Solids: Optical Properties of Electrons in Matter* (Cambridge University Press, 2002).

[65] L.K. Dash, F. Bruneval, V. Trinite, N. Vast, and L. Reining, Comp. Mater. Sci. **38**, 482 (2007).

[66] V. Mauchamp, F. Boucher, and P. Moreau, Ionics **14**, 191 (2008).

[67] K.W.K. Shung, Phys. Rev. B **34**, 979 (1986).

[68] V. Mauchamp, P. Moreau, G. Ouvrard, and F. Boucher, Phys. Rev. B **77**, 045117 (2008).

[69] K. Andersen, K.W. Jacobsen, and K.S. Thygesen, Phys. Rev. B **86**, 245129 (2012).

[70] X. Rocquefelte, F. Goubin, Y. Montardi, N. Viadere, A. Demourgues, A. Tressaud, M.-H. Whangbo, and S. Jobic, Inorg. Chem. **44**, 3589 (2005).

[71] A. Marinopoulos, L. Reining, V. Olevano, A. Rubio, T. Pichler, X. Liu, M. Knupfer, and J. Fink, Phys. Rev. Lett. **89**, 076402 (2002).

[72] M. Bujoli-Doeuff, R. Dessapt, P. Deniard, and S. Jobic, Inorg. Chem. **51**, 142 (2012).

[73] R. Dessapt, D. Kervern, M. Bujoli-Doeuff, P. Deniard, M. Evain, and S. Jobic, Inorg. Chem. **49**, 11309 (2010).

**TAB. I:** Static optical dielectric constant, Re[$\varepsilon(\omega=0)$], calculated along the **x**, **y** and **z** directions with and without local-field effects. The directions of the Cartesians projector, **x**, **y** and **z,** follow the same directions as the direct lattice vectors, **a**, **b**, and **c**.

|              | x   | y   | z   | Average |
|--------------|-----|-----|-----|---------|
| **Without LFE** | 7.2 | 6.8 | 6.5 | 7.0     |
| **With LFE**    | 5.5 | 4.2 | 6.3 | 5.3     |

**FIG. 1:** (Color Online) (a) Ball and stick representation of orthorhombic α-MoO$_3$. The $^2/_\infty$[MoO$_3$] sheets are stacked along the [010] direction, and are held together via Van der Waals interactions (grey sphere: molybdenum, orange sphere: oxygen). (b) Schematic building of the $^2/_\infty$[MoO$_3$] layer in α-MoO$_3$ from the linear $^1/_\infty$[MoO$_4$] chain displayed along two directions perpendicular to the running axis. The $^1/_\infty$[MoO$_4$] chains are connected via corner sharing to give rise to the $^2/_\infty$[MoO$_3$] sheet (blue octahedra: [MoO$_6$]).

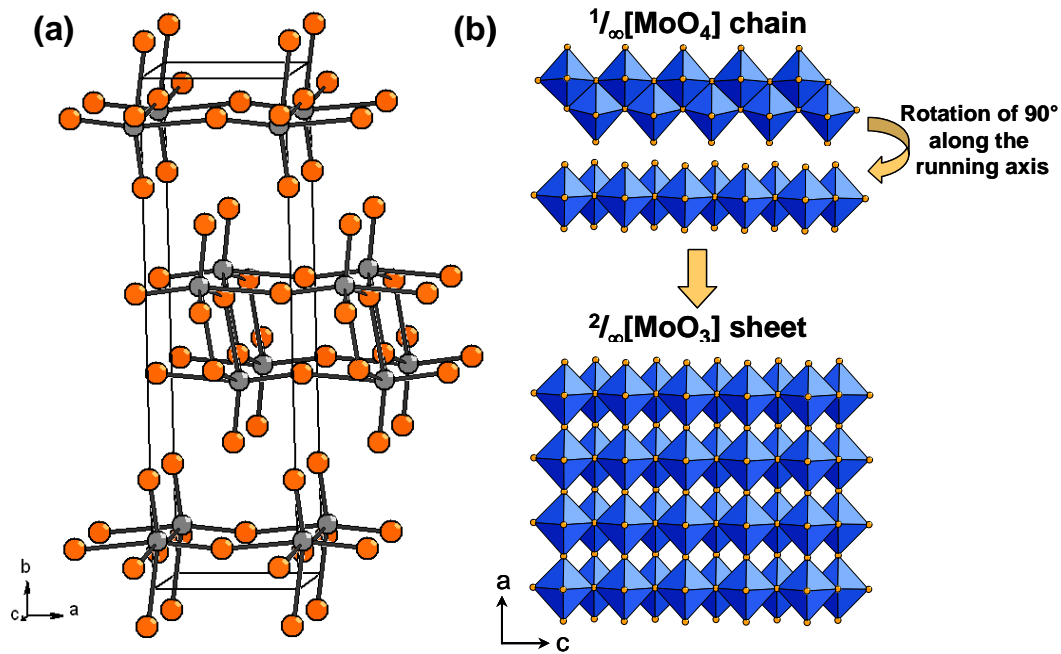

**FIG. 2:** (Color Online) Comparison of the experimental VEELS spectrum of α-MoO$_3$ recorded along the [210] zone axis (thick black line) with the calculated ones including or not local-field effects (LFE): thin red line and thin dotted line, respectively.

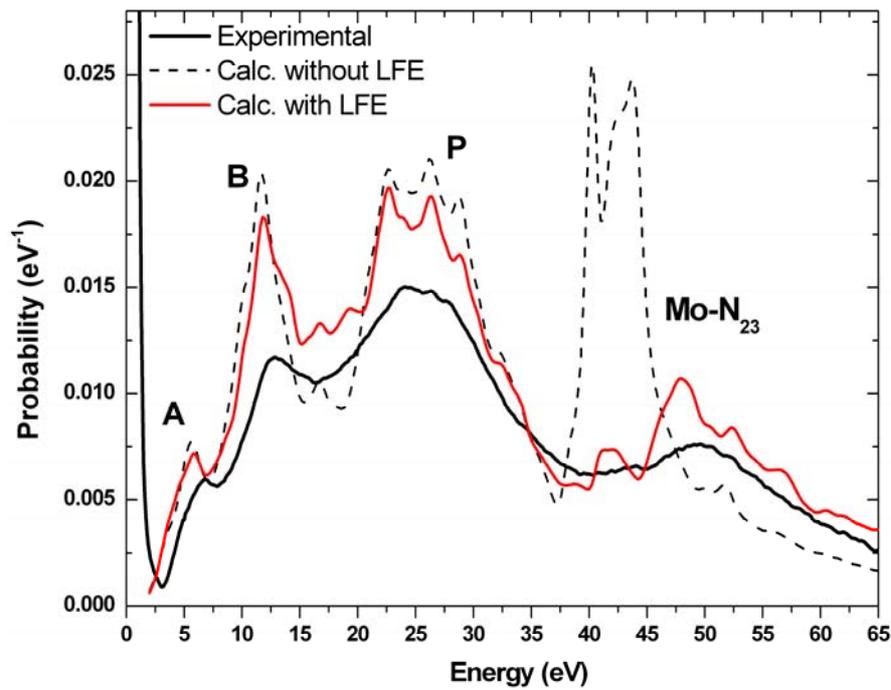

**FIG. 3:** (Color Online) Comparison of energy-loss functions (ELF), including or not LFE (thick red line and thin black line, respectively), calculated along the *x* (a), *y* (b) and *z* (c) axes.

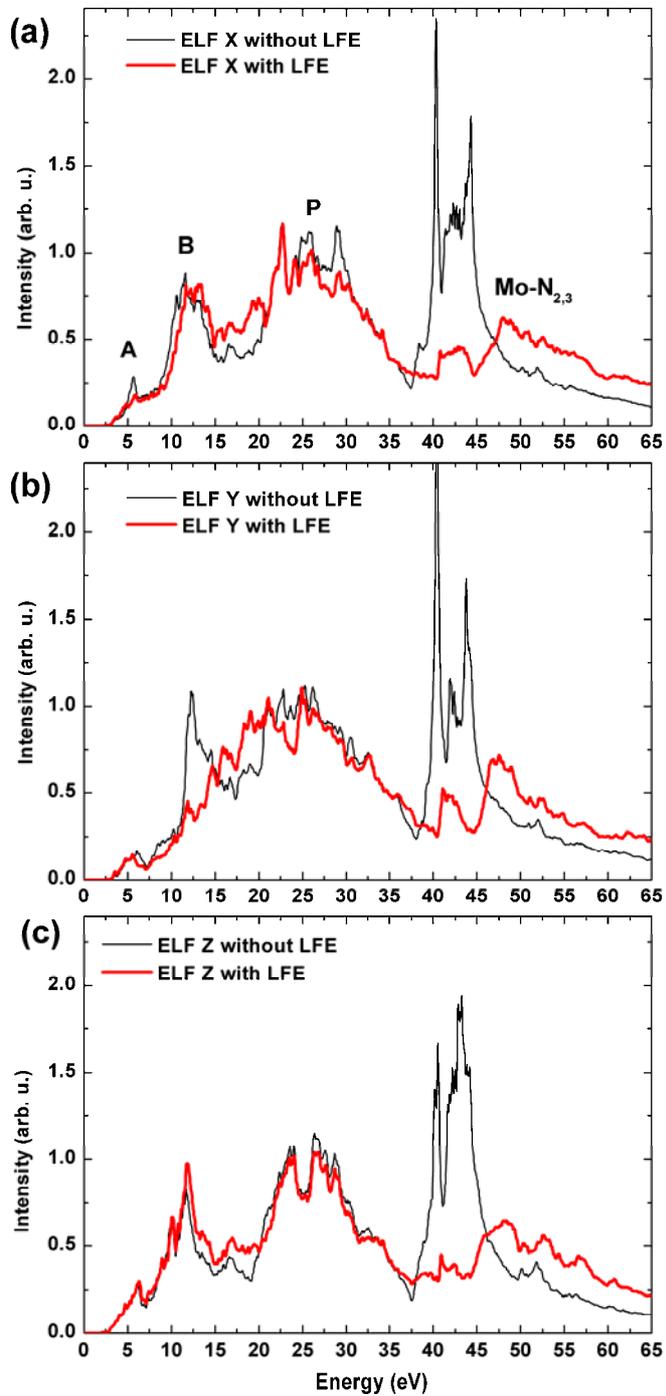

**FIG. 4:** (Color Online) Calculated energy-loss function (ELF, thick black line) together with the imaginary ($\varepsilon_2$, red line) and real ($\varepsilon_1$, blue dots) parts of $\varepsilon(\mathbf{0},\omega)$ without (a) and with (b) LFE calculated along the *x* axis.

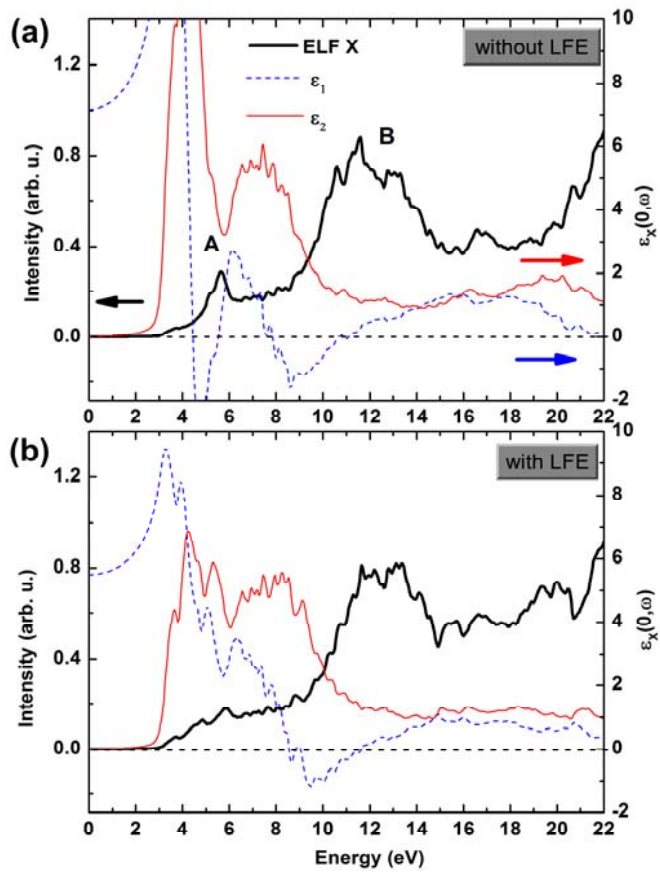

**FIG. 5:** (Color Online) calculated energy-loss function (ELF, thick black line) together with the imaginary ($\varepsilon_2$, red line) and real ($\varepsilon_1$, blue dots) parts of $\varepsilon(\mathbf{0},\omega)$ without (a) and with (b) LFE calculated along the *y* axis.

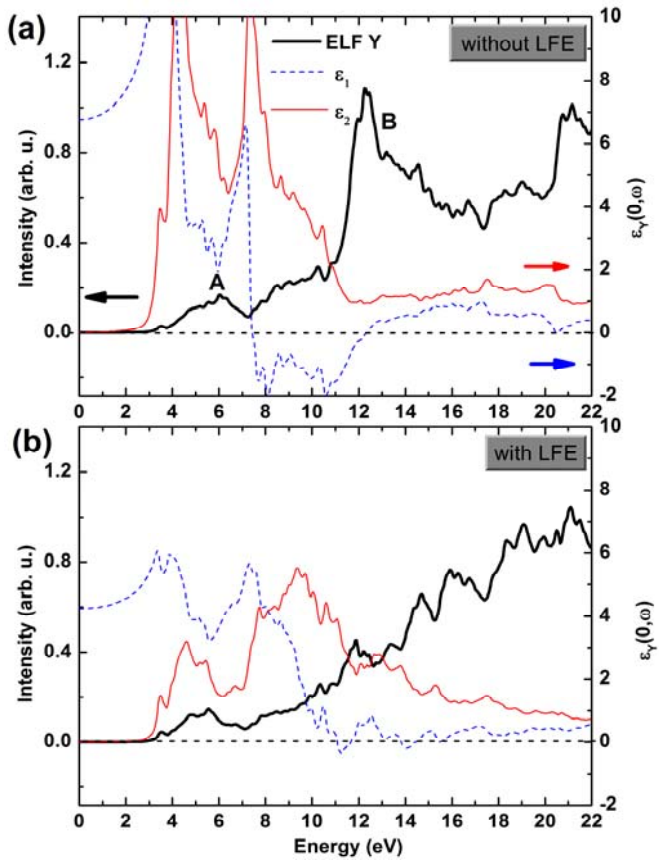

**FIG. 6:** (Color Online) Imaginary part of $\varepsilon_y(\mathbf{0},\omega)$ calculated by considering only the transitions from the 53 to 56$^{th}$ bands to the 77 and 78$^{th}$ bands (thick red line). For comparison the result obtained considering all the transitions is also given (thin black line).

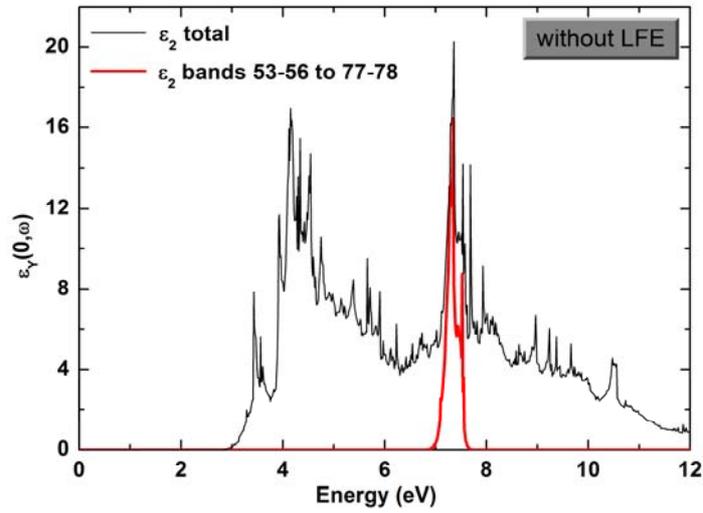

**FIG. 7:** (Color Online) Kohn-Sham band structure of α-MoO$_3$. The conduction bands 77 and 78 involved in the 7 eV peak are plotted in green. The O-$p_y$ character of the bands 53-56 is highlighted in red circles and in cyan for the other valence bands.

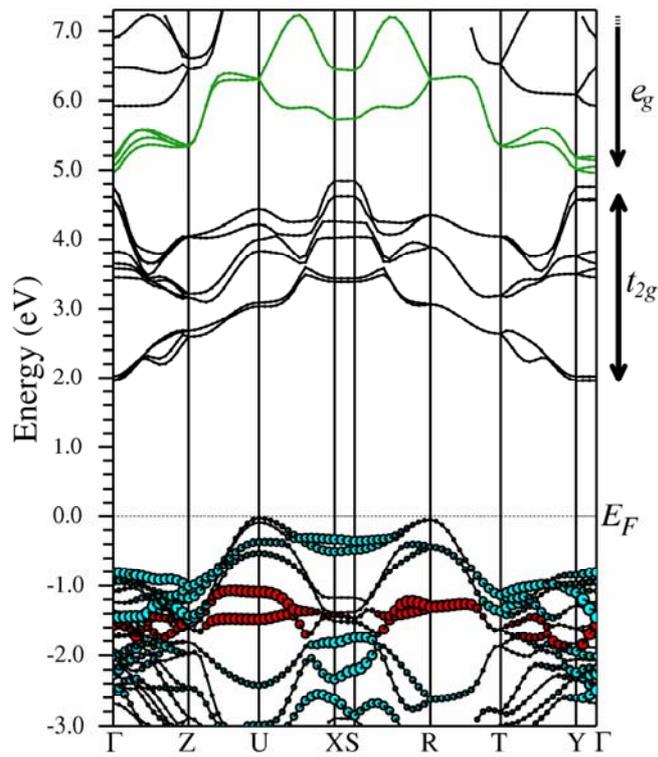

**FIG. 8:** (Color Online) Atomic orbital decomposition of the wave functions at the $k$-point $L_1$ when considering only the bands 53-56 and the bands 77-78.

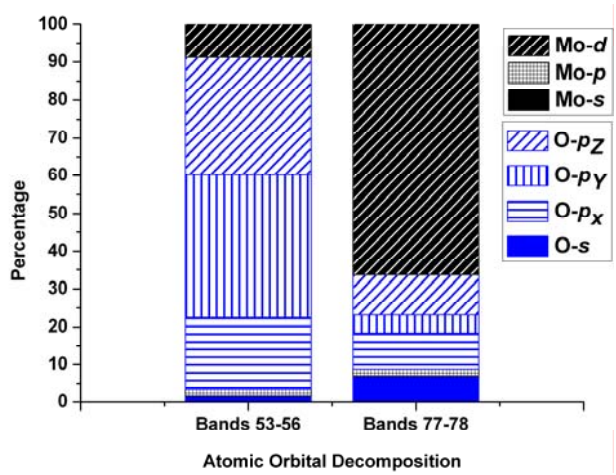

**FIG. 9:** (Color Online) Electron density like representation (square modulus of the wave function) at the *k*-point $L_1$ for the bands 53 -56 (a) and the bands 77-78 (b). The red circles highlight the depletion of the electron density around the terminal oxygens (grey sphere: molybdenum, orange sphere: oxygen).

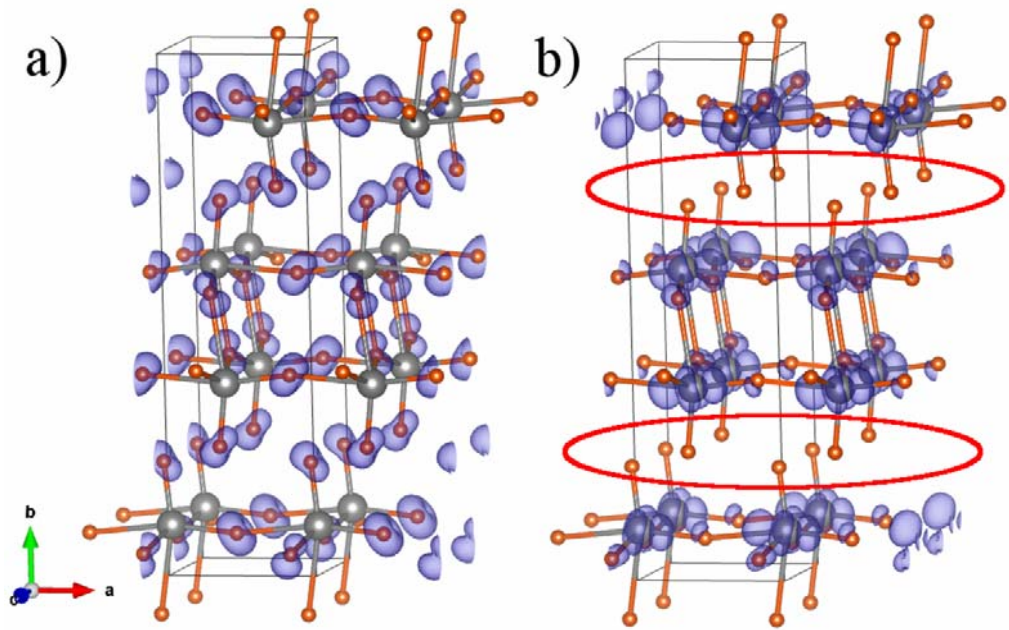

**FIG. 10:** (Color Online) Imaginary part of ε(**0**,ω) calculated along the *x* (a) and *y* (b) axes for α-MoO$_3$ (red lines) and MoO$_3$ with an increased interlayer spacing of 2 Å (black lines). For each structure and each direction, the results calculated without LFE (dashed lines) and with LFE (continuous lines) are compared.

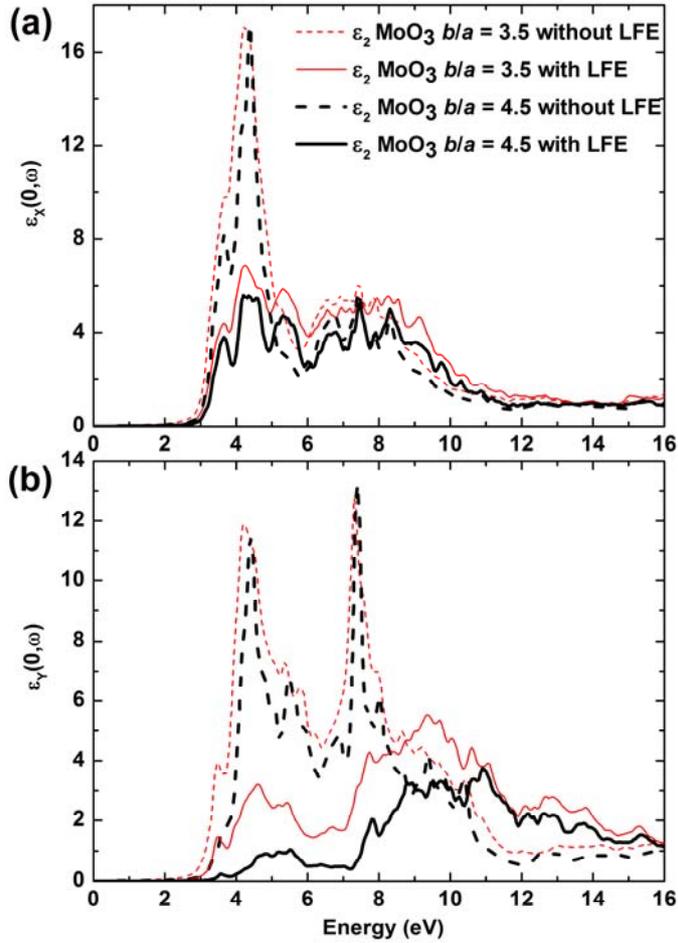

Supplemental Material for

Strong anisotropic influence of local-field effects on the dielectric response of α-MoO$_3$

*L. Lajaunie*, F. Boucher, R. Dessapt, and P. Moreau.*

Institut des Matériaux Jean Rouxel, (IMN) – Université de Nantes, CNRS, 2 rue de la Houssinère - BP 32229, 44322 Nantes Cedex 3, France.

* Corresponding author: luc.lajaunie@cnrs-imn.fr

**FIG. 1:** (Color Online) MoO$_6$ octahedra highlighting the bond lengths and bond angles notations. Structurally the oxygen sites can be divided into three categories: terminal (atom labeled 1), twofold coordinated (2 and 3) and threefold coordinated (4 and 5).

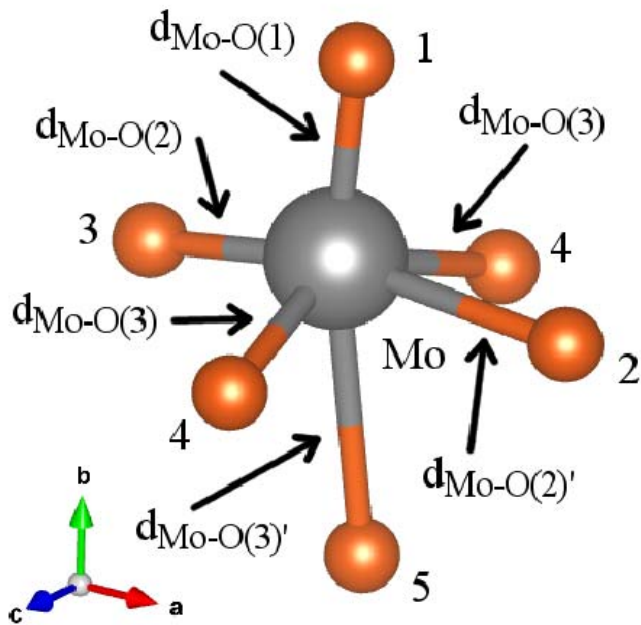

**TAB. 1:** Calculated lattice constants and bond lengths (in Å) of α-MoO$_3$ compared with previous experimental and theoretical works.

|  | APO* | FO** | PBE-GGA^ [58] | PBE+DFT-D2 [57] | Expt. [1] |
|---|---|---|---|---|---|
| **a** | 3.9624 | 3.908 | 4.023 | 3.931 | 3.9624(1) |
| **b** | 13.860 | 13.831 | 13.855 | 13.881 | 13.860(2) |
| **c** | 3.6971 | 3.694 | 3.755 | 3.711 | 3.6971(4) |
| **d$_{Mo-O(1)}$** | 1.70 | 1.68 | 1.70 | 1.70 | 1.68 |
| **d$_{Mo-O(2)}$** | 1.76 | 1.76 | 1.76 | 1.77 | 1.74 |
| **d$_{Mo-O(2)'}$** | 2.22 | 2.18 | 2.28 | 2.19 | 2.24 |
| **d$_{Mo-O(3)}$** | 1.95 | 1.94 | 1.98 | 1.95 | 1.95 |
| **d$_{Mo-O(3)'}$** | 2.40 | 2.41 | 2.34 | 2.40 | 2.31 |

\* Atomic Positions Optimization (PBE-GGA)
\*\* Full Optimization (PBE-GGA+DFT-D2)
^ Parameter *b* fixed

**TAB. 2:** Calculated bond angles (°) compared with previous experimental work.

|  | APO | FO | Expt.[1] |
|---|---|---|---|
| **1-Mo-2** | 89.7 | 91.0 | 87.8 |
| **1-Mo-3** | 102.8 | 103.0 | 102.7 |
| **1-Mo-4** | 105.0 | 104.8 | 104.2 |
| **1-Mo-5** | 166.2 | 167.4 | 164.7 |
| **2-Mo-3** | 167.5 | 166.1 | 169.5 |
| **2-Mo-4** | 79.1 | 79.6 | 79.0 |
| **2-Mo-5** | 76.5 | 76.4 | 76.9 |
| **3-Mo-4** | 97.4 | 96.6 | 98.1 |
| **3-Mo-5** | 91.0 | 89.8 | 92.6 |
| **4-Mo-4** | 142.5 | 143.9 | 143.1 |
| **4-Mo-5** | 72.7 | 73.3 | 73.2 |

In Table 1, structural parameters obtained with APO (PBE-GGA) and FO (PBE-GGA+DFT-D2) methods are compared. The overall agreement with the experimental data[1] is excellent for both approaches. In particular concerning the lattice constants, semi-empirical DFT-D2 correction leads to an excellent agreement with experiment,[1] the calculated lattice constants being only slightly reduced: 1.4% for *a*, 0.2% for *b* and 0.3% for *c*. These results are very similar to those obtained from previous calculation using the same approach[57] and constitute a clear improvement over previous PBE-GGA calculations[58] where the lattice parameter *b* has to be rescaled when the relaxation of cell parameters is concerned. Considering now bond lengths, results obtained by APO and FO methods show an excellent agreement with available experimental data and previous calculations with the exception of vdW-corrected calculations of $d_{Mo-O(2)'}$ and $d_{Mo-O(3)'}$ (~ 2.6 and 4.8% smaller and larger than experimental data, respectively). Considering the bond angles, they all present an excellent agreement with the experimental data. Finally, whatever the calculation method used (APO or FO), the dielectric properties up to 80 eV are found to be identical (not shown here). Thus, structural parameters obtained from the APO method have been used to calculate the dielectric properties.

**FIG. 2:** (Color Online) Calculated energy-loss function (ELF, thick black line) together with the imaginary ($\varepsilon_2$, red line) and real ($\varepsilon_1$, blue dots) parts of $\varepsilon(\mathbf{0},\omega)$ without (a) and with (b) LFE calculated along the $z$ axis.

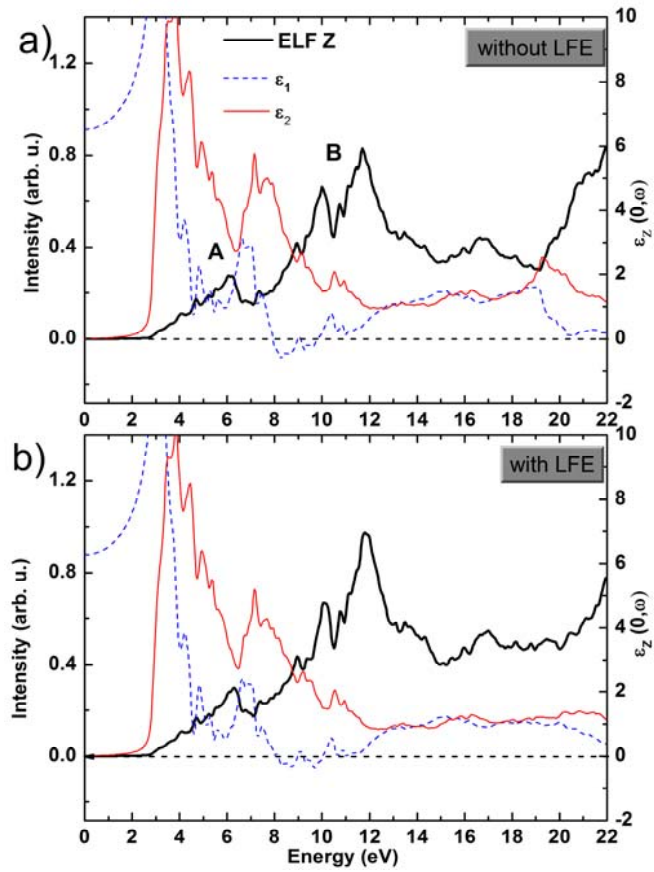